\documentclass[letterpaper, 10 pt, conference]{ieeeconf}
\IEEEoverridecommandlockouts 
\usepackage{graphicx} 
\usepackage{cite}
\usepackage{caption}
\usepackage{amsmath}
\usepackage{amsfonts}
\usepackage{dsfont}
\let\labelindent\relax
\usepackage{enumitem,kantlipsum}
\usepackage{subcaption}

\date{} 

\title{\LARGE \bf Adaptive Traffic Light Control for Competing Vehicle and Pedestrian Flows}

\author{Yingqing Chen and Christos G. Cassandras
\thanks{Y. Chen and C. G. Cassandras are
with the Division of Systems Engineering and Center for Information and
Systems Engineering, Boston University, Brookline, MA 02446
\tt\small\{yqchenn;cgc\}@bu.edu.}
}

\begin{document}
\maketitle
\thispagestyle{empty}
\pagestyle{empty}

\begin{abstract} 
We study the Traffic Light Control (TLC) problem for a single intersection, considering both straight driving vehicle flows and corresponding crossing pedestrian flows with the goal of achieving a fair jointly optimal sharing policy in terms of average waiting times.
Using a stochastic hybrid system model, we design a quasi-dynamic policy controlling the traffic light cycles with several threshold parameters applied to the light cycles and the partially observed contents of vehicle and pedestrian queues. Infinitesimal Perturbation Analysis (IPA) is then used to derive a data-driven gradient estimator of a cost metric with respect to the policy parameters and to iteratively adjust these parameters through an online gradient-based algorithm in order to improve overall performance on this intersection and adapt the policy to changing traffic conditions. The controller is applied to a simulated intersection in the town of Veberöd, Sweden, to illustrate the performance of this approach using real traffic data from this intersection.  
\end{abstract}



\section{INTRODUCTION}
The Traffic Light Control (TLC) problem entails dynamically adjusting the traffic light cycles in an intersection or a set of intersections in order to improve the overall traffic performance (normally measured through a congestion metric). Different traffic models for intersections may be formulated (such as in \cite{WeyWann-Ming2000Mfas} and \cite{van2007integrated}), which provide the basis for optimization. With models of different levels of detail, methods such as optimization of a mixed integer quadratic programming problem (MIQP) in \cite{zhang_traffic_2017}, model predictive control (MPC) in \cite{lin_efficient_2012}, and a linear-quadratic regulator (LQR)in \cite{wang_optimizing_2022} are widely used. Computational intelligence methods and techniques have also been applied to the TLC problem, including artificial neural networks (ANNs), fuzzy systems, evolutionary computation (EC) algorithms, and reinforcement learning (RL) \cite{dongbin_zhao_computational_2012}. For example, \cite{bi_optimal_2018} developed a two-layer type-2 fuzzy controller which can not only improve the traffic situation of each intersection but also consider downstream intersections and enlarge the so-called ``green wave'' band. In \cite{dong_analysis_2019} an adaptive fuzzy neural network algorithm is used to learn from the historical data and make online adjustments, while \cite{kaur_adaptive_2014} used a Genetic Algorithm to adapt the traffic signal time. Moreover, \cite{chu_multi-agent_2020} developed a form of decentralized multi-agent RL algorithm that can be applied to large-scale TLC problems showing the capability of achieving lower and more sustainable intersection delays, by distributing the traffic more homogeneously among intersections.

Research to date has focused on vehicle flows, while the presence of pedestrians has been largely ignored, even though it is clear that it plays an  important role. In the current research literature, there is limited work on how pedestrians influence traffic at an intersection.  In \cite{ishaque_multimodal_1939}, a simulation study was conducted using VISSIM to determine the effects of signal cycle timings on the delay caused to both vehicles and pedestrians and to come to an optimal signal cycle under different demand levels. While such simulations are set with a fixed preset traffic light pattern, \cite{ma_optimization_2015} focused more on the pedestrian phase pattern and established a detailed pedestrian delay model to compare both safety and efficiency between an exclusive pedestrian phase (EPP) and a normal two-way crossing (TWC) pattern. Also considering EPP, \cite{gomes_traffic_2022} built a mathematical optimization model to optimize the allocation of times for traffic lights placed at intersections for a relatively static traffic scenario, and presented a meta-heuristic algorithm to solve the problem. Similarly, \cite{zhang_traffic_2017} modeled both vehicle and pedestrian flows based on a cell transmission model and translated it into a mixed integer quadratic programming (MIQP) problem with the objective of minimizing the weighted sum of pedestrian and vehicle delay time. Illegal crossing behaviors by pedestrians have also been considered in \cite{zhang_pedestrian-safety-aware_2021} which presented a traffic signal control strategy for improving vehicle passing efficiency as well as pedestrian crossing safety. A Genetic Algorithm (GA) and a Harmony Search (HS) algorithm are used to solve this problem.

Although these methods show good performance, most of them need to train the controller with a large amount of historical data or need a prohibitive amount of computation for a single static traffic scenario (with computational complexity rapidly increasing in more crowded situations).  
With recent technological developments that allow the real-time detection of vehicles and pedestrians at intersections
(e.g., \cite{sreekumar_real-time_2018}), it has recently become possible to develop real-time traffic-responsive strategies
instead of solely relying on historical traffic data.
In the context of such traffic-responsive TLC policies, Infinitesimal Perturbation Analysis (IPA) \cite{cassandras_perturbation_2010}
provides data-driven unbiased gradient estimates of performance metrics with respect to various system design or control parameters
from a single observed sample path, i.e., current traffic information. IPA has been used in TLC problems, e.g., \cite{panayiotou_online_2005} used the IPA gradients of the queue length with respect to the green/red light lengths within a signal cycle with fixed cycle length, while \cite{fleck_adaptive_2016} relaxed the cycle length constraint and developed a quasi-dynamic control framework, improving the overall system performance of a single intersection by simultaneously adjusting queue content thresholds and green cycle length thresholds. Beyond a single intersection, IPA was used in \cite{geng_multi-intersection_2012} to derive TLC policies for a multi-intersection problem.

In this paper, we study the TLC problem for a single intersection modeled as a stochastic hybrid system where the traffic light switching process is event-driven, while the dynamics of the vehicle and pedestrian flows through an intersection are time-driven. This gives rise to a a Stochastic Flow Model (SFM) as in \cite{fleck_adaptive_2016}. IPA is used to estimate online gradients of a performance metric with respect to several parameters of a quasi-dynamic TLC policy 
considering both vehicle flows and pedestrian flows. These gradient estimates are then used 
to iteratively seek optimal values for these system parameters. Compared to \cite{fleck_adaptive_2016}, the presence of pedestrian flows requires the SFM to include additional queues and the TLC policy to incorporate conditions for enabling pedestrian flows to cross, thus creating a new trade-off between vehicle and pedestrian performance metrics.

The remainder of this paper is organized as follows. In section 2, we formulate the TLC problem for a single intersection and present the SFM modeling framework. Section 3 details the derivation of the IPA estimators for a cost function gradient with respect to a controllable parameter vector. The IPA estimators are then incorporated into a gradient-based optimization algorithm. In Section 4, we conduct simulation experiments based on the real traffic situations at a specific intersection in the town of Veberöd, Sweden, and show results supporting the effectiveness of our TLC IPA-driven policy. Finally, we conclude and discuss future work in Section 5.

\section{Problem Formulation}\label{sec:problem}

\subsection{Signalized Intersection Modeling}

\begin{figure}
\begin{center}
\includegraphics[width=5cm]{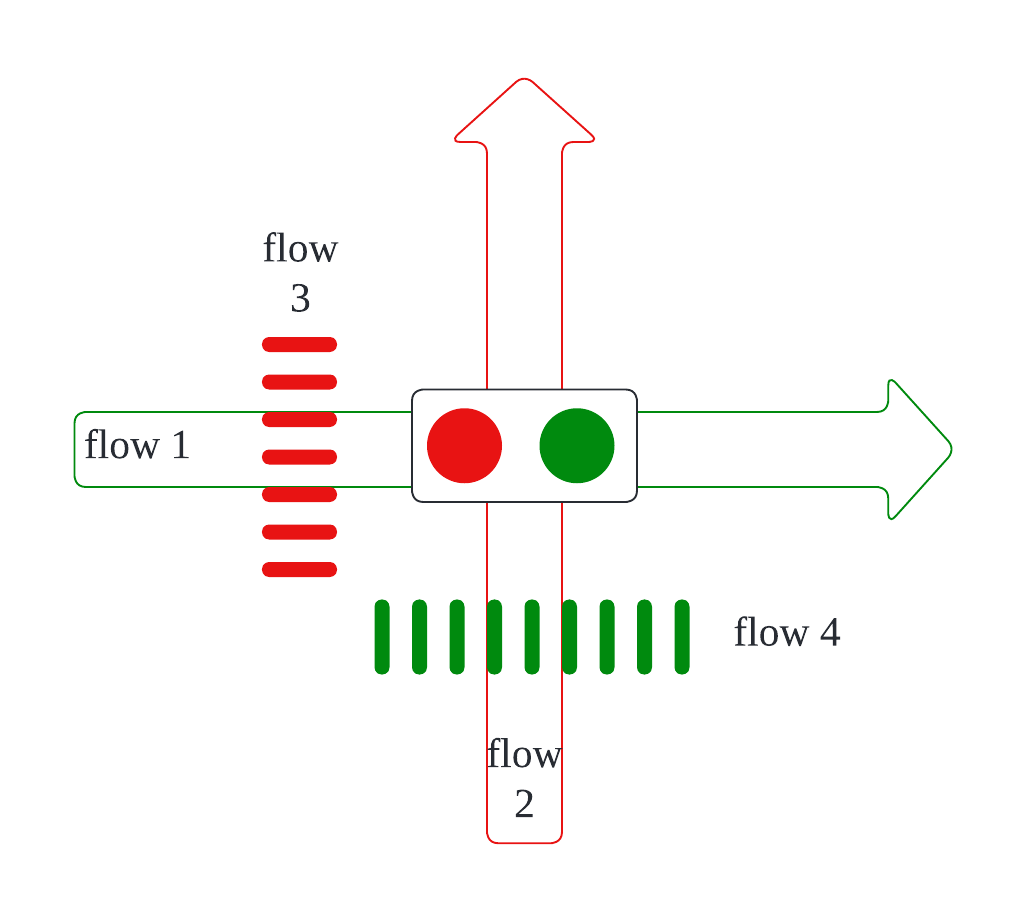}
\caption{\label{fig1}Single intersection with two roads and pedestrian crossings}
\end{center}
\end{figure}

Consider a single signalized intersection as shown in Fig. \ref{fig1}. For simplicity, left-turn and right-turn traffic flows are not considered, and the traffic light combines yellow with red. We consider only two vehicle flows (perpendicular to each other) indexed by $n=1,2$ and two corresponding pedestrian flows indexed by $n=3,4$ (see Fig. \ref{fig1}).
A basic requirement for TLC is that the signals for vehicles and pedestrians are consistent, i.e., when vehicles of flow 1(2) face a GREEN light, pedestrians of flow 4(3) must also face a GREEN light, as indicated in Fig. \ref{fig1}.

Each of the two roads is modeled as a queue where vehicles may stop when facing a RED light. Similarly, each sidewalk where pedestrians wait to cross a road is modeled as a queue. Thus, we define a state vector $x(t)=[x_1(t), x_2(t), x_3(t), x_4(t)]$, $x_n(t) \in \mathbb{R}_0^+$ of flow content queues
corresponding to the four flows. Similar to \cite{fleck_adaptive_2016}, we model the input to each queue as an exogenous random flow process
$\{ \alpha_n(t) \}$ where $\alpha_n(t)$ is a stochastic instantaneous arrival rate independent of all queue states.
When the traffic light corresponding to queue $n$ is GREEN, the departure flow process is denoted by 
$\{ \beta_n(t) \}$. 
Note that when $n=1,2$, $x_n(t)$ denotes the vehicle queue content of road $n$, while when $n=3,4$, $x_n(t)$ represents the associated pedestrian queue content intending to cross road $n-2$. 

In addition, we define a clock state variable $z_n(t)$ for $n=1,2$ to measure the time since the last switch from RED to GREEN for vehicle flow 1,2 respectively. Henceforth, let $\bar{n}$ denote the index of the vehicle flow perpendicular to flow $n$ with the requirement that $z_{\bar n} (t)>0$ when $z_n(t)=0$, for all $t\ge0$.
Similarly, for the pedestrian flows, we define $w_n(t)$, $n=3,4$ to be the time elapsed since the presence of the first pedestrian in queue $n$ in the current RED phase;
in other words, $w_n(t)$ captures the longest pedestrian waiting time in this phase. 

Letting $z(t)=[z_1(t), z_2(t)]$ and $w(t)=[w_3(t), w_4(t)]$ with $z_n(t), w_n(t) \in \mathbb{R}_0^+$, we have the 8-dimensional system state vector $[x(t), z(t), w(t)]$. Before presenting the detailed state dynamics, we define the traffic light controller:
\begin{align} \label{udef}
    u(x(t), z(t), w(t))=[u_1(t),u_2(t),u_3(t), u_4(t)]
\end{align}
where $u_n(t)=1$ denoting a GREEN light faced by flow $n$, and $u_n(t)=0$ denoting a RED light accordingly. We define $u_n(t)$ to be right-continuous in order to accurately represent the control policy defined in the sequel. Due to the basic safety constraint already mentioned, i.e., when vehicles of flow 1(2) face a GREEN light, pedestrians of flow 4(3) must also face a GREEN light,
we can eliminate the control values that would lead to a flow conflict. As a result, the feasible control set contains only two elements: $U=\{[1,0,0,1],[0,1,1,0]\}$ and anytime $u_n(t)$ switches its value, the control for the other three flows switches automatically. 

Since $\alpha_n(t)$ is an exogenous input process independent of the queue states, we can write the departure process as:
\begin{align} \label{beta}
  \beta_n(t) =
    \begin{cases}
      h_n(t), & \text{if $x_n(t)>0$ and $u_n(t)=1$}\\
      \alpha_n(t), & \text{if $x_n(t)=0$ and $u_n(t)=1$}\\
      0, & \text{otherwise}
    \end{cases}   
\end{align}
for $n=1,2,3,4$, where $h_n(t)$ is the maximum departure rate which generally depends on the road structure, vehicle specifications and pedestrian move pattern. For the pedestrian flows $n=3,4$, we assume $h_n(t)>\alpha_n(t)$ for all $t$, i.e., once pedestrians start crossing, the queue definitely gets shorter.

We can now write the state dynamics as follows:
\begin{equation}
\label{eqn:xdynamic}
  \dot x_n(t) =\alpha_n(t)-\beta_n(t), ~~~~~~~~~~~~~n=1,2,3,4
\end{equation}
\begin{equation}
\label{eqn:zdynmic12}
  \dot z_n(t) =
    \begin{cases}
      1, & \text{if $u_n(t)=1$ }\\
      0, & \text{otherwise} ~~~~~~~~~~~~~~n=1,2
    \end{cases}       
\end{equation}
\begin{equation}
\label{eqn:zdynmic34}
  \dot w_n(t) =
    \begin{cases}
      1, & \text{if $u_n(t)=0$ and $x_{n}(t)>0$ }\\
      0, & \text{otherwise} ~~~~~~~~~~~~~~n=3,4
    \end{cases}       
\end{equation}
In (\ref{eqn:zdynmic12}), note that $\dot z_1(t) + \dot z_2(t)=1$ always holds. In addition, we define $z_n(t)$ to be left-continuous, so that at the moment the light switches from GREEN to RED, $z_n(t)>0$,  $z_{\bar{n}}(t)=0$, $z_n(t^+)=0$, and $u_n(t)=0$ (since $u_n(t)$ is right-continuous). 
In (\ref{eqn:zdynmic34}), note that $w_n(t^+)=0$ whenever $u_n(t)$ switches from 0 to 1, and $w_3(t)w_4(t)=0$ always holds.

Thus, the traffic light intersection in Fig. \ref{fig1} can be viewed as a hybrid system in which the time-driven dynamics are
given by (\ref{eqn:xdynamic}), (\ref{eqn:zdynmic12}), (\ref{eqn:zdynmic34}) and (\ref{beta}), while event-driven dynamics are
associated with light switches with events that cause the value of $x_n(t)$ to change from strictly positive to zero or
vice versa. Although the dynamics are based on knowledge of the instantaneous flow processes $\{ \alpha_n(t) \}$ and $\{ \beta_n(t) \}$, we will show that the IPA-based adaptive controller we design \emph{does not require such knowledge} and depends only on estimating such rates in the vicinity of certain critical observable events.

\subsection{Controller Specification}\label{section:Controller Specification}
\begin{figure}[ht]
\begin{subfigure}{.26\textwidth}
  \centering
  \includegraphics[width=4.6cm]{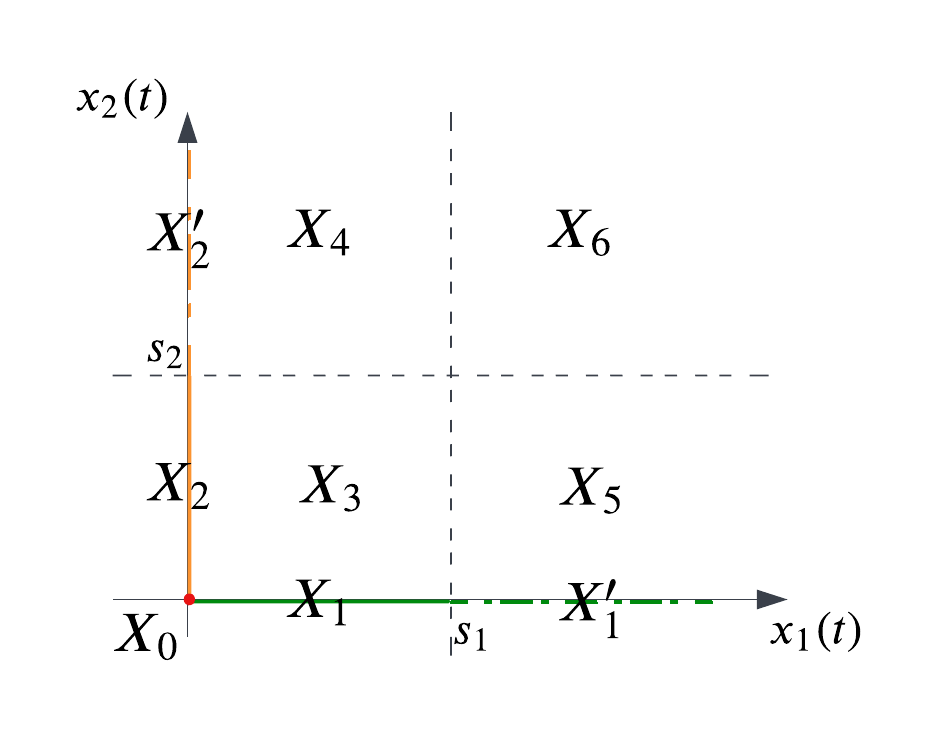}  
  \caption{State-space partition for vehicle queue content}
  \label{fig:state_rep_vehicle}
\end{subfigure}
\begin{subfigure}{.22\textwidth}
  \centering
  \includegraphics[width=4.0cm]{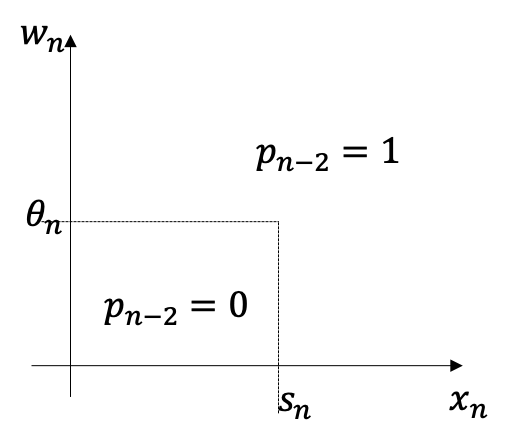}  
  \caption{State-space partition for pedestrian state $p_n(t),~ (n=3,4)$}
  \label{fig:state_rep_p}
\end{subfigure}
\caption{State space partitions}
\label{fig:state_rep}
\end{figure}

Our TLC design for the intersection in Fig. \ref{fig1} is based  on the ability of current sensors to detect events of interest in the state dynamics above, such as a queue content becoming empty. While it may not be possible to detect the exact number of vehicles in a queue (e.g., using cameras), we assume that this number can be estimated so as to classify a queue content $x_n(t)$ as being empty and either below or above some threshold $s_n$, $n=1,2,3,4$, as well as the time such transitions occur. For the vehicle queue contents, the joint state space can be partitioned into the following nine regions (as shown in Fig. \ref{fig:state_rep_vehicle}):\\
$ X_0=\{(x_1,x_2):x_1(t)=0, x_2(t)=0 \} $\\
$ X_1=\{(x_1,x_2):0<x_1(t)<s_1, x_2(t)=0 \} $ \\
$ X_1'=\{(x_1,x_2):x_1(t)\geq s_1, x_2(t)=0 \} $ \\
$ X_2=\{(x_1,x_2):x_1(t)=0, 0<x_2(t)<s_2 \} $\\
$ X_2'=\{(x_1,x_2):x_1(t)=0, x_2(t)\geq s_2 \} $\\
$X_3=\{(x_1,x_2):0<x_1(t)<s_1, 0<x_2(t)<s_2 \}$\\
$X_4=\{(x_1,x_2):0<x_1(t)<s_1, x_2(t)\geq s_2 \}$\\
$X_5=\{(x_1,x_2):x_1(t)\geq s_1, 0<x_2(t)<s_2 \}$\\
$X_6=\{(x_1,x_2):x_1(t)\geq s_1, x_2(t)\geq s_2 \}$

Regarding the length of a light cycle (i.e., the values that $z_n(t)$, $n=1,2$, can take),
we assign a guaranteed minimum GREEN light cycle time $\theta_n^{min}$ and a maximum cycle time $\theta_n^{max}$. This is to ensure that traffic light switches are not overly frequent nor can they be excessively long. In fact, the adaptivity of our controller largely rests on its ability to adjust on line the parameters $\theta_n^{min}$, $\theta_n^{max}$, in addition to a few others defined next, based on observed events (fully defined in the sequel) and their occurrence times. Complementing $\theta_n^{max}$, $n=1,2$, for vehicles, we define an upper bound $\theta_n$ to the pedestrian waiting times $w_n(t)$, $n=3,4$, so that their waiting never becomes excessive.

An efficient controller design also needs to address the issues of maintaining $(i)$ a proper balance between allocating a GREEN light to competing queues and $(ii)$ preventing the undesired phenomenon where vehicles wait at a RED light at road $n$ while road $\bar{n}$ is empty during its GREEN phase. Such ``waiting-for-nothing'' instances waste the resources of vehicles that wait unnecessarily and can be eliminated through a proper controller design as detailed next. Towards these two goals, the final parameters we define for our TLC design are the queue thresholds $s_n$, $n=1,2,3,4$ (see Fig. 2). To summarize, we define the following controllable parameter vector:
\begin{equation} \label{parametervector}
    \upsilon = [\theta_1^{min}, \theta_1^{max}, \theta_2^{min}, \theta_2^{max}, \theta_3, \theta_4, s_1,s_2,s_3,s_4]
\end{equation}
where $\theta_n^{min} \geq 0$, $\theta_n^{max} \geq \theta_n^{min}$, and $\theta_{n+2}>0$ for $n=1,2$; $s_n>0$ for $n=1,2,3,4$. 


The role of these controllable parameters, is to partition the 8-dimensional state space into appropriate subsets that form the basis of a \emph{quasi-dynamic} controller: while in (\ref{udef}) the controller is defined as a function of the full state $[x(t), z(t), w(t)]$, a quasi-dynamic controller is a function of subsets of aggregated states defined by the partition of the queue states shown in 
Fig. \ref{fig:state_rep_vehicle} and an additional partition shown in Fig. \ref{fig:state_rep_p}; the latter is based on defining the following auxiliary state variable for each pedestrians flow $n=3,4$:  
\begin{equation}
\label{eqn:p definition}
  p_{n-2}(x_n(t), w_n(t)) =
    \begin{cases}
      1, & \text{if $x_{n}(t) \geq s_{n}$ OR $w_{n}(t) \geq \theta_{n}$}\\
      0, & \text{otherwise}
    \end{cases}       
\end{equation}
We use the simplified notation $p_{n-2}(t)$ that captures when pedestrian queue $n$ is enabled to cross: when the queue is either long enough or the waiting time is large enough during the current RED phase (see Fig. \ref{fig:state_rep_p}). The subscript $n-2$ indicates the target road for pedestrian queue $n$ to cross. Thus, we write the aggregated pedestrian state as $p(t)=[p_1(t),p_2(t)]$. 


We are now ready to specify a quasi-dynamic controller expressed as $u(X(t), p(t), z(t), w(t))$ where $X(t)$ is one of the nine subsets defined by the partition of the vehicle queue content state space in Fig. \ref{fig:state_rep_vehicle} and $p(t)$ is the aggregated pedestrian state. 
Due to the coupling between vehicle and pedestrian demands, a control policy can no longer be as simple as the one as in \cite{fleck_adaptive_2016} where no pedestrian flows were considered. With the goal of balancing access to a GREEN light to ensure fairness and preventing the wasteful
``waiting-for-nothing'' phenomenon mentioned earlier, we express the TLC specification as conditions for either maintaining a GREEN light or switching back to it for road 1. Recall that $U=\{[1,0,0,1],[0,1,1,0]\}$, hence the policy for $u_1(t)$ fully defines $u_2(t), u_3(t), u_4(t)$.

\begin{itemize}[wide]
    \item [1.] $(x_1(t), x_2(t)) \in \{X_0\}$:
In this case there are no vehicle queues and control is only applied to serve pedestrian queues. First, if the light in road 1 is GREEN ($z_1>0$), it should remain GREEN as long as pedestrian demand for crossing road 1 is low ($p_1=0$). It switches to RED only when it reaches its upper bound ($\theta_1^{max}$)
as long as pedestrian demand is high for both roads ($p_1=p_2=1$). Second, if the light in road 1 is RED ($z_2>0$), the logic for switching to GREEN is symmetric. Formally, we define  
\begin{equation}
\label{eqn:control rule X0}
  u_1(t) =
    \begin{cases}
      1, & \text{if [$z_1(t)\in (0, \theta_1^{max})$, $p_1(t)=p_2(t)=1$] OR}\\
      & \text{ [$z_1(t)>0$, $p_1(t)=0$] OR}\\
      & \text{ [$z_2(t)\geq \theta_2^{max}$, $p_1(t)=p_2(t)=1$] OR}\\
        & \text{ [$z_2(t)>0$, $p_1(t)=0, p_2(t)=1$] }\\
      0, & \text{otherwise}
    \end{cases}       
\end{equation}

\item [2.] $(x_1(t), x_2(t)) \in \{X_1, X_1'\}$:
In this case there is no vehicle flow in road 2, so control is applied to serve the pedestrian queue that may form in road 1.
First, if the light in road 1 is GREEN ($z_1>0$), it remains GREEN for at least $\theta_1^{min}$ to serve the vehicles, and then switches only if unbalanced pedestrian demand for road 1 arises ($p_1(t) > p_2(t)$). Second, if the light in road 1 is RED ($z_2>0$), it switches back to GREEN as soon as pedestrian demand for road 1 is low or the GREEN cycle for road 2 has reached its upper bound $\theta_2^{max}$.

\begin{align}
\label{eqn:control rule X1}
  u_1(t) =
    \begin{cases}
      1, & \text{if [$z_1(t)\in (0, \theta_1^{min})$] OR} \\
      & \text{[$z_1(t)\geq \theta_1^{min}$, $p_1(t)\leq p_2(t)$] OR}\\
      & \text{[$z_2(t)\in (0,\theta_2^{max})$, $p_1(t)=0$] OR} \\
      & \text{[$z_2(t)\geq \theta_2^{max}$]} \\
      0, & \text{otherwise}
    \end{cases}       
\end{align}

\item[3. ]$(x_1(t), x_2(t)) \in \{X_2, X_2'\}$:
In this case there is no vehicle flow in road 1 so control is applied to serve the pedestrian queue that may form in road 2.
The control logic is symmetric to Case 2.
\begin{equation}
\label{eqn:control rule X2}
  u_1(t) =
    \begin{cases}
      1, & \text{if [$z_1(t)\in (0, \theta_1^{max})$, $p_2(t)=1$] OR} \\
      & \text{[$z_2(t)\geq \theta_2^{min}$, $p_1(t)=0, p_2(t)=1$]} \\
      0, & \text{otherwise}
    \end{cases}        
\end{equation}

\item[4. ] $(x_1(t), x_2(t)) \in \{X_3, X_6\}$:
In this case, vehicle traffic is balanced for the two roads (either low or high as in Fig. \ref{fig:state_rep_vehicle}). First, if the light in road 1 is GREEN ($z_1>0$), it remains GREEN for at least $\theta_1^{min}$ and then then switches only if unbalanced pedestrian demand for road 1 arises ($p_1(t) > p_2(t)$). Second, if the light in road 1 is RED ($z_2>0$), the logic for maintaining or switching to GREEN is symmetric.
\begin{equation}
\label{eqn:control rule X36}
  u_1(t) = 
    \begin{cases}
      1, & \text{if [$z_1(t)\in (0, \theta_1^{min})$] OR} \\
      & \text{[$z_1(t)\in [\theta_1^{min}, \theta_1^{max})$, $p_1(t)\leq p_2(t)$] OR}\\
      & \text{[$z_2(t) \in [\theta_2^{min}, \theta_2^{max})$, $p_1(t)=0, p_2(t)=1$]} \\
      & \text{ OR [$z_2(t)\geq \theta_2^{max}$]}\\
      0, & \text{otherwise}
    \end{cases}       
\end{equation}

\item[5. ] $(x_1(t), x_2(t)) \in \{X_4\}$:
In this case, there exists low vehicle traffic demand for road 1 and high vehicle traffic demand for road 2, thus vehicle crossing should be prioritized. First, if the light in road 1 is GREEN ($z_1>0$), it remains GREEN only until it reaches its lower bound ($\theta_1^{min}$). Second, if the light in road 1 is RED ($z_2>0$), it remains RED until the road 2 reaches its GREEN cycle upper bound $\theta_2^{max}$ in order to cope with the heavy road 2 traffic.

\begin{equation}
\label{eqn:control rule X4}
  u_1(t) =
    \begin{cases}
      1, & \text{if [$z_1(t)\in (0, \theta_1^{min})$] OR} \\
      & \text{[$z_2(t)\geq \theta_2^{max}$$]$} \\
      0, & \text{otherwise}
    \end{cases}       
\end{equation}

\item[6. ] $(x_1(t), x_2(t)) \in \{X_5\}$:
In this case, there is high vehicle traffic demand for road 1 and low vehicle traffic demand for road 2, so that light switching logic is symmetric to Case 5.

\begin{equation}
\label{eqn:control rule X5}
  u_1(t) =
    \begin{cases}
      1, & \text{if [$z_1(t)\in (0, \theta_1^{max})$] OR} \\
      & \text{[$z_2(t)\geq \theta_2^{min}$]} \\
      0, & \text{otherwise}
    \end{cases} 
\end{equation}

\end{itemize}
The six cases above fully specify our TLC. It's important to note that as the values of $x_n(t)$ change for $n=1,2$, there are associated transitions from one subset in
Fig. \ref{fig:state_rep_vehicle}) to another, in which case the controller $u_1(t)$ follows the new region specifications.

\subsection{Event definitions}\label{event definitions}

We begin by defining all observable events associated with mode switches in the hybrid system defined through 
(\ref{eqn:xdynamic}), (\ref{eqn:zdynmic12}), (\ref{eqn:zdynmic34}) and (\ref{beta}) under the controller specified above.

For $n=1,2,3,4$: (1) $x_n$ reaches 0 from above ($x_n\downarrow 0$),
(2) $x_n$ becomes positive from 0 ($x_n\uparrow 0$),
(3) $x_n$ reaches $s_n$ from below ($x_n\uparrow s_n$),
(4) $x_n$ reaches $s_n$ from above ($x_n\downarrow s_n$),
(5) $\alpha_n$ reaches 0 from above ($\alpha_n\downarrow 0$),
(6) $\alpha_n$ becomes positive from 0 ($\alpha_n\uparrow 0$).

For $n=1,2$: (1) $z_n$ reaches lower bound ($z_n\uparrow \theta_n^{min}$),
(2) $z_n$ reaches upper bound ($z_n\uparrow \theta_n^{max}$).

For $n=3,4$: $w_n$ reaches threshold ($w_n\uparrow \theta_n$).

Since we have defined the auxiliary state variables $p_n(t)$, it is convenient to also define the following compound events for $n=1,2$:
(1) $p_n$ changes from 0 to 1 ($p_n\uparrow 1$), which happens when either $x_{n+2}\uparrow s_{n+2}$ or $w_{n+2}\uparrow \theta_{n+2}$ occurs,
(2) $p_n$ changes from 1 to 0 ($p_n\downarrow 0$), which happens only at a GREEN phase when $x_{n+2}\downarrow s_{n+2}$ occurs.

We can now define the \emph{controllable} event $G2R_n$ $n=1,2,3,4$, which switches the light faced by queue $n$ from GREEN to RED 
and triggers a state mode in the hybrid system to switch. This is the event that causes a control switch from $u_n(t)=1$ to $u_n(t)=0$. 
Similarly, $R2G_n$ indicates light switches from RED to GREEN. 
Note that these controllable events are coupled, i.e., when $G2R_1$ occurs, then $R2G_2$, $R2G_3$, $G2R_4$ also occur at the same time (see Fig. \ref{fig1}).

\begin{figure*}
\begin{center}
\captionsetup{justification=centering}
\includegraphics[width=1.0\textwidth]{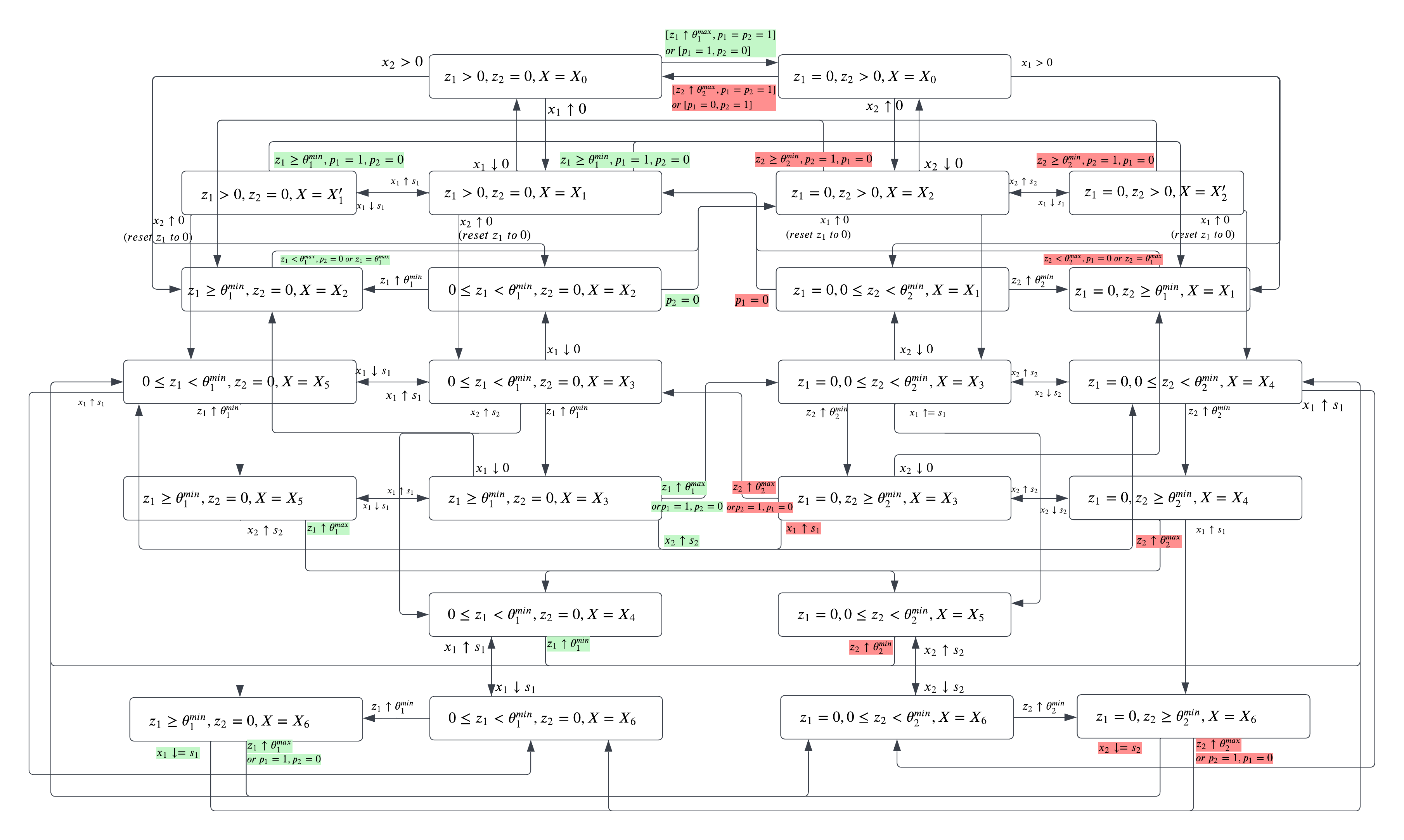}
\caption{\label{fig3}SHA under quasi-dynamic control}

\end{center}
\end{figure*}

It is important to observe that all $G2R_n$ controllable events are fully defined through the observable events defined above, and based on the logic defined in section \ref{section:Controller Specification}.
As an example, 
when $(x_1(t), x_2(t)) \in \{X_4\}$, $G2R_1$ occurs when [$z_1\uparrow \theta_1^{min}$] according to (\ref{eqn:control rule X4}). Additionally, when [$x_1\downarrow 0$] happens, $X(t)$ jumps from $X_4$ to $X_2'$ and then follow the logic of (\ref{eqn:control rule X2}). So that when $p_2(t)=0$ at the event time, $G2R_1$ occurs too. 
Finally, a complete representation of the hybrid system under TLC is provided in Fig. \ref{fig3} in the form of a Stochastic Hybrid Automaton (SHA) model. In the figure, events with a green background are those leading to $G2R_1$, while the events with red background are events leading to $R2G_1$.




\subsection{TLC Optimization Problem}

\begin{figure}
\begin{center}
\includegraphics[width=7.4cm]{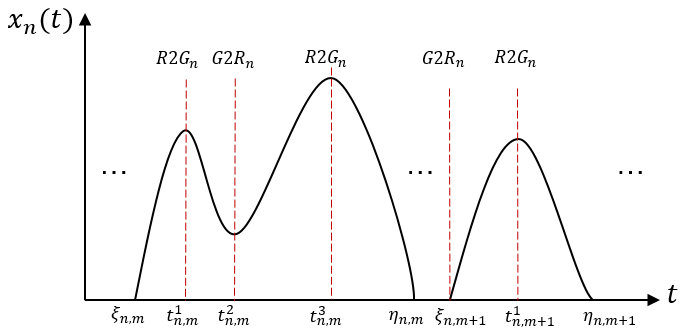}
\caption{\label{fig4}Typical sample path of a traffic light queue}
\end{center}
\end{figure}

With the parameterized controller defined above, our aim is to optimize a performance metric for the intersection operation with respect to these controllable parameters that comprise the vector $\upsilon$ defined in (\ref{parametervector}). 
We choose our performance metric to be the weighted mean of all queue lengths over
a fixed time interval $[0,T]$:
\begin{equation} \label{Lfunction}
    L(\upsilon; x(0), z(0), T) = \frac{1}{T}\sum_{n=1}^{4}\int_{0}^{T} \omega_n x_n(\upsilon,t) \,dt 
\end{equation}
where $\omega_n$ is a weight associated with queue $n$, $n=1,2,3,4$. In order to focus on the structure of a typical sample path of the hybrid system modeling the intersection under TLC, note that such a sample path of the flow queue content $\{x_n(t)\}$ consists of alternating Non-empty Periods (NEPs) and Empty Periods(EPs), which correspond to time intervals when $x_n(t)>0$ and $x_n(t)=0$ respectively, as shown in Fig. \ref{fig4}. The sample path includes light switching events ($R2G_n$ or $G2R_n$), events starting NEPs (denoted as $S_n$), and events starting EPs (denoted as $E_n$) all of which are associated with the observable events defined earlier. Moreover, we denote the $m$th NEP of queue $n$ by $[\xi_{n,m}, \eta_{n,m})$ where $\xi_{n,m}$, $\eta_{n,m}$ are the occurrence times of the $m$th $S_n$ event and $m$th $E_n$ event respectively. 
Since $x_n(t)=0$ during EPs of queue $n$, the sample function $L(\upsilon; x(0), z(0), T)$ in (\ref{Lfunction}) can be rewritten as
\begin{equation}
\label{eqn:L}
    L(\upsilon; x(0), z(0), T) = \frac{1}{T}\sum_{n=1}^{4}\sum_{m=1}^{M_n}\int_{\xi_{n,m}}^{\eta_{n,m}} \omega_n x_n(\upsilon,t) \,dt 
\end{equation}
where $M_n$ is the total number of NEPs during the sample path of queue $n$ over $[0,T]$,
and $\xi_{n,m}$, $\eta_{n,m}$ are the start and end time of the $m$th NEP respectively.

Thus, our goal is to determine $\upsilon$ that minimizes the expected weighted mean queue length:
\begin{equation}
    J(\upsilon; x(0), z(0), T) = E[L(\upsilon; x(0), z(0), T)]
\end{equation}
We note that it is not possible to derive a closed-form expression of $J(\upsilon; x(0), z(0), T)$ even if we had full knowledge of the processes 
$\{ \alpha_n(t) \}$ and $\{ \beta_n(t) \}$. Therefore, a closed-form expression for $\nabla J(\upsilon)$ is also infeasible if a gradient-based method is invoked to solve this minimization problem. 
The role of IPA is to obtain an unbiased estimate of $\nabla J(\upsilon)$ based
on the sample function gradient $\nabla L(\upsilon)$ which can be evaluated based
only on data directly observable along a single sample path such as Fig. \ref{fig4}, as will be shown in the next section.
The unbiasedness of $\nabla L(\upsilon)$ is ensured
under mild conditions on $L(\upsilon)$ (see \cite{cassandras_perturbation_2010}) and assuming that $\alpha_n(t)$ and $h_n(t)$ are piecewise continuously differentiable in $t$ w.p. 1. In particular, we emphasize that \emph{no explicit knowledge of $\alpha_n(t)$ and $h_n(t)$ is necessary} to estimate $\nabla J(\upsilon)$ through $\nabla L(\upsilon)$.

We can now invoke a gradient-based algorithm of the form
\begin{equation} \label{gradient_algo}
    \upsilon_{i,l+1} = \upsilon_{i,l}-\rho_l \big[\frac{dJ}{d\upsilon_{i,l}}\big]_{IPA}
\end{equation}
where $\upsilon_{i,l}$ is the $i$th parameter of $\upsilon$ at the $l$th iteration, $\rho_l$ is the stepsize at the $l$th iteration, and $(\frac{dJ}{d\upsilon_{i,l}})_{IPA}$ is the IPA estimator of $\frac{dJ}{d\upsilon_{i,l}}$, which will be derived in the next section.



\section{Infinitesimal Perturbation Analysis}
We begin with a brief review of the IPA framework in \cite{cassandras_perturbation_2010}. Consider a sample path over $[0,T]$ and denote the occurrence time of the $k$th event (of any type) by $\tau_k$. Let $x'(\theta, t)$, $\tau'_k(\theta
)$ be the derivatives of $x(\theta, t)$, $\tau_k(\theta)$ over the scalar controllable parameter of interest $\theta$ respectively. We omit the dependence on $\theta$ for ease of notation hereafter. The dynamics of $x(t)$ are fixed over any interevent interval$[\tau_k, \tau_{k+1})$, represented by $\dot{x}(t)=f_k(t)$. Then, the state derivative satisfies 
\begin{equation}
\label{eqn:state_derivative_pre}
    \frac{d}{dt}x'(t)=\frac{\partial f_k(t)}{\partial x}x'(t) + \frac{\partial f_k(t)}{\partial \theta}
\end{equation}
with boundary condition:
\begin{equation}
\label{eqn:state_derivative}
    x'(\tau_k^+) = x'(\tau_k^-) + [f_{k-1}(\tau_k^-)-f_k(\tau_k^+)]\tau_k'
\end{equation}

In order to evaluate (\ref{eqn:state_derivative}), $\tau_k'$ should be determined, which depends on the type of event taking place at $\tau_k$.

For exogenous events (events causing a discrete state transition that is independent of any controllable parameter):
\begin{equation}
    \tau'_{k,i}=0, ~~i=1,...10, ~~k=1,2,...
\end{equation}
For endogenous events (events that occur when there exists a continuously differentiable function $g_k$ such that $\tau_k=min\{t>\tau_{k-1}:g_k(x(\theta, t),\theta)=0\}$) with guard condition $g_k = 0$:
\begin{equation}
\label{eqn:event_time_derivative}
    \tau_k' = -[\frac{\partial g_k}{\partial x} f_k(\tau_k^-)]^{-1}(\frac{\partial g_k}{\partial \theta}+\frac{\partial g_k}{\partial x}x'(\tau_k^-))
\end{equation}
This framework captures how system states and event times change with respect to controllable parameters. Our goal is to estimate $\nabla J(\upsilon)$ through $\nabla L(\upsilon)$, and, according to (\ref{eqn:L}), the performance metric expression is a function of event time and system state variables. Thus, we apply the IPA framework to the TLC problem and evaluate how a perturbation in $\upsilon$ would affect performance metrics.

\subsection{State Derivatives}
We define the derivatives of the state variables $x_n(t)$, $z_n(t)$, $w_n(t)$ and event time $\tau_k$ with respect to parameter $\upsilon_i$ ($i=1,...10$) as follows:
\begin{equation}
     x_{n,i}'\equiv \frac{\partial x_n(t)}{\partial \upsilon_i}, 
     z_{n,i}'\equiv \frac{\partial z_n(t)}{\partial \upsilon_i},  
    w_{n,i}'\equiv \frac{\partial w_n(t)}{\partial \upsilon_i},  
   \tau_{k,i}'\equiv \frac{\partial \tau_k}{\partial \upsilon_i},  
\end{equation}
Also, we denote the state dynamics at interval time $t\in [\tau_k, \tau_{k+1})$ as follows:
\begin{equation*}
    \dot x_n(t) = f_{n,k}^x(t),  n = 1,2,3,4
\end{equation*}
\begin{equation*}
    \dot z_n(t) = f_{n,k}^z(t), n=1,2
\end{equation*}
\begin{equation}
    \dot w_n(t) = f_{n,k}^w(t), n=3,4
\end{equation}
Using the dynamics in (\ref{eqn:xdynamic}), (\ref{eqn:zdynmic12}) and (\ref{eqn:zdynmic34}) in (\ref{eqn:state_derivative_pre}), similar to the analysis in \cite{fleck_adaptive_2016},
we can easily conclude that the state derivative of any queue would not change within a mode, i.e., for $t\in[\tau_k,\tau_{k+1})$:
\begin{equation}\label{eqn:derivative within mode}
    x'_{n,i}(t)=x'_{n,i}(\tau_k^+),
    z'_{n,i}(t)=z'_{n,i}(\tau_k^+),
    w'_{n,i}(t)=w'_{n,i}(\tau_k^+)
\end{equation}

For any event time $\tau_k$, we partition the associated events into the following subsets and determine the value of all queue content derivatives when an event occurs. For any flow $n=1,2,3,4$ and any controllable parameter $\upsilon_i$, $i=1,...10$:
\begin{itemize}
    \item[1)] Events inside an EP: Since $x_n(t)=0$ throughout an EP, it immediately follows that
    \begin{equation}
    \label{eqn:state_derivative_EP}
        x_{n,i}'(\tau_k^+) = 0 
    \end{equation}
    
    \item[2)]Events that start an EP at queue $n$ (denoted by $E_n$): This is when flow $n$ is facing a GREEN and an endogenous event [$x_n\downarrow 0$] takes place. The state dynamics are $f_{n,k-1}^x(\tau_k^-)=\alpha_n(\tau_k)-h_n(\tau_k)$, $f_{n,k}^x(\tau_k^+)=0$. Then from (\ref{eqn:state_derivative}), for any $i = 1,...10$:
    \begin{equation}
    \label{eqn:state_derivative_En}
        x_{n,i}'(\tau_k^+)=x_{n,i}'(\tau_k^-)+(\alpha_n(\tau_k)-h_n(\tau_k))\tau_{k,i}'
    \end{equation}
    
    \item[3)] Events that start a NEP at queue $n$ (denoted by $S_n$): there are two ways in which this event may occur:
    \begin{itemize}
        \item[3.1)] $S_n$ is induced by a light switching event, specifically a $G2R_n$ event. We have $f_{n,k-1}^x(\tau_k^-)=0$ and $f_{n,k}^x(\tau_k^+)=\alpha_n(\tau_k)$. Based on (\ref{eqn:state_derivative}), for any $i = 1,...10$:
        \begin{equation}
        \label{eqn:state_derivative_Sn_G2R}
            x_{n,i}'(\tau_k^+)=-\alpha_n(\tau_k)\tau'_{k,i}.
        \end{equation}
        \item [3.2)]$S_n$ is not related to light switching, i.e, the NEP starts when queue $n$ faces either GREEN or RED because an exogenous change in $\alpha_n(t)$ takes place. Since this event is exogenous,  $\tau_k'=0$. Thus, based on (\ref{eqn:state_derivative}), for any $i=1,...10$:
        \begin{equation}
        \label{eqn:state_derivative_Sn_no_switch}
            x_{n,i}'(\tau_k^+) = x_{n,i}'(\tau_k^-)=0
        \end{equation}
    \end{itemize}
    
    \item[4)] Events inside a NEP but not induced by light switching. It is obvious that in this case $f_{n,k-1}^x(\tau_k^-)=f_{n,k}^x(\tau_k^+)$, so that $x_{n,i}'(\tau_k^+) = x_{n,i}'(\tau_k^-)$ from (\ref{eqn:state_derivative}).
    
    \item [5)]Events inside a NEP and induced by light switching. These events possess the most information affecting the system state and can be further classified as follows:
    \begin{itemize}
        \item[5.1)] $G2R_n$. It follows from (\ref{eqn:xdynamic}) that $f_{n,k-1}^x(\tau_k^-)=\alpha_n(\tau_k)-h_n(\tau_k)$ and $f_{n,k}^x(\tau_k^+)=\alpha_n(\tau_k)$. Then, from (\ref{eqn:state_derivative}), for any $i=1,..10$:
        \begin{equation}
        \label{eqn:state_derivative_NEP_G2R}
            x'_{n,i}(\tau_k^+)=x'_{n,i}(\tau_k^-)-h_n(\tau_k)\tau'_{k,i}
        \end{equation}
        \item [5.2)]$R2G_n$. It follows from (\ref{eqn:xdynamic}) that $f_{n,k-1}^x(\tau_k^-)=\alpha_n(\tau_k)$ and $f_{n,k}^x(\tau_k^+)=\alpha_n(\tau_k)-h_n(\tau_k)$. Then, for any $i=1,...10$:
        \begin{equation}
        \label{eqn:state_derivative_NEP_R2G}
            x'_{n,i}(\tau_k^+)=x'_{n,i}(\tau_k^-)+h_n(\tau_k)\tau'_{k,i}
        \end{equation}
    \end{itemize}
\end{itemize}

Note that (\ref{eqn:state_derivative_En}),(\ref{eqn:state_derivative_Sn_G2R}),(\ref{eqn:state_derivative_NEP_G2R}) and (\ref{eqn:state_derivative_NEP_R2G}) all depend on the event time derivatives $\tau_{k,i}'$ which we derive next.

\subsection{Event Time Derivatives}
The determination of $\tau_{k,i}'$ is based on applying (\ref{eqn:event_time_derivative}) for the specific events and dynamics in our system.
The event time derivatives of interest are shown below in two groups by their range of application. The notation $\mathds{1}_{i=m}$ denotes the indicator function whose value is $1$ when $i=m$.
The detailed analysis can be found in the Appendix.

For $n=1,2$:
\begin{equation}
\label{eqn:event_derivative_total1}
    \tau_{k,i}' =
    \begin{cases}
    
      \mathds{1}_{i=2n}+\tau_{k-1,i}', & \text{if [$z_n\uparrow \theta_n^{max}$] occurs at $\tau_k$}\\
      \mathds{1}_{i=2n-1}+\tau_{k-1,i}', & \text{if [$z_n\uparrow \theta_n^{min}$] occurs at $\tau_k$}\\ 
      \mathds{1}_{i=n+4}, & \text{if [$w_{n+2}\uparrow \theta_{n+2}$] occurs at $\tau_k$}\\
    \end{cases} 
\end{equation}

For $n=1,\ldots,4$:
\begin{equation}
\label{eqn:event_derivative_total2}
    \tau_{k,i}' =
    \begin{cases}
      \frac{\mathds{1}_{i=n+6}-x'_{n,i}(\tau_k^-)}{\alpha_n(\tau_k)-h_n(\tau_k)}, & \text{if [$x_n\downarrow s_n$] occurs at $\tau_k$}\\
      \frac{\mathds{1}_{i=n+6}-x'_{n,i}(\tau_k^-)}{\alpha_n(\tau_k)}, & \text{if [$x_n\uparrow s_n$] occurs at $\tau_k$}\\
        \frac{-x_{n,i}'(\tau_k^-)}{\alpha_n(\tau_k)-h_n(\tau_k)} & \text{if [$x_n\downarrow 0$] occurs at $\tau_k$}\\
        0, & \text{if [$\alpha_n\downarrow 0$] occurs at $\tau_k$}\\
    \end{cases} 
\end{equation}
where all events were defined in Section \ref{event definitions} and $i=1,\ldots,10$.  

\subsection{Cost Derivatives}
With the state and event time derivatives obtained from an observed sample path of the SFM, we can now derive IPA cost derivative from (\ref{eqn:L}). Similar to the proof in \cite{fleck_adaptive_2016}, the IPA estimator, i.e., the derivative of $L(\upsilon)$ with respect to $\upsilon_i$, $i=1\ldots10$, is given by
\begin{equation}
\label{eqn:ipa1}
    \frac{dL(\upsilon)}{d\upsilon_i} = \frac{1}{T}\sum_{n=1}^{4}\sum_{m=1}^{M_n}\omega_n\frac{dL_{n,m}(\upsilon)}{d\upsilon_i}
\end{equation}
where
\begin{multline}
\label{eqn:ipa2}
    \frac{dL_{n,m}(\upsilon)}{d\upsilon_i}=x_{n,i}'({\xi_{n,m}}^+)(t_{n,m}^1-\xi_{n,m})\\
    +x_{n,i}'({t_{n,m}^{J_{n,m}}}^+)(\eta_{n,m}-t_{n,m}^{J_{n,m}})\\
    +\sum_{j=2}^{J_{n,m}}x_{n,i}'({t_{n,m}^j}^+)(t_{n,m}^j-t_{n,m}^{j-1})
\end{multline}
where $J_{n,m}$ is the total number of events related to queue $n$ in the $m$th NEP, $t_{n,m}^j$ is the time of the $j$th event in that NEP, and $\xi_{n,m}$, $\eta_{n,m}$ are the start and end time respectively of of $m$th NEP. 

It is clear from the three terms in (\ref{eqn:ipa2}) that the IPA derivative is the sum of certain inter-event times multiplied by their corresponding state derivatives. Therefore, the information required to evaluate it consists of: event times, which are easy to observe; 
state derivatives at these times, which can be obtained from (\ref{eqn:state_derivative_EP}) through (\ref{eqn:ipa2});
and arrival and departure rates $\alpha_n(\tau_k)$, $h_n(\tau_k)$, needed \emph{only} certain event times (e.g., when event $S_n$ is induced by light switching in (\ref{eqn:state_derivative_Sn_G2R})). The later are easy to estimate, as detailed in the next section. Moreover, we can assume the maximum departure rate $h_n(\tau_k)$ to be a constant which can also be easily estimated offline. In summary, by simply monitoring and recording 
events as they are observed and very limited calculations, we can obtain the IPA gradient estimator for the expected weighted mean queue length. This is then used with any standard online gradient-based algorithm (\ref{gradient_algo}) so as to adjust the controllable parameters and improve the overall performance.

\section{Simulation Results}
\label{sec:sim}
In this section, we use Eclipse SUMO (Simulation of Urban MObility) to build a simulation environment for traffic through a single traffic light intersection. 
Although the arrival processes can be arbitrary for our IPA-based method, we assume them to be Poisson processes for both vehicles and pedestrians with rates $\bar{\alpha}_n, n=1,\ldots,4$, and estimate the maximum departure rate as a constant value $h_n=H, n=1,\ldots,4$ through an offline analysis.
Since we only need the arrival flow rate at certain event times, we can estimate an instantaneous arrival rate through $\alpha_n(\tau_k)=N_a/t_w$, where $N_a$ denotes the number of vehicles/pedestrians joining queue $n$ during a time window of size $t_w$ before event time $\tau_k$; this is easy to detect and record in SUMO.  We set $H=1.2$ and equal weight for all flows ($w_n=1$) throughout this section. With this setting, we have performed four sets of simulations: one for an intersection where we test the controller's ability to optimize our cost metric, one applied to real-world intersection with observed traffic statistics, one for testing the online implementation, and the last one to test the adaptivity of the controller. 

\begin{figure*}[ht]
\begin{subfigure}{.24\textwidth}
  \centering
  \includegraphics[width=4cm]{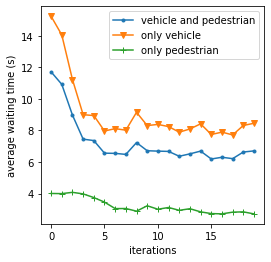} 
  \caption{}
  \label{fig:c_performance}
\end{subfigure}
\begin{subfigure}{.24\textwidth}
  \centering
  \includegraphics[width=4.0cm]{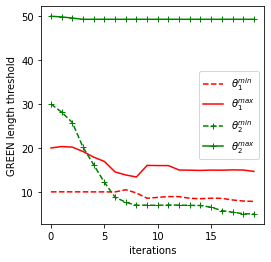}  
  \caption{}
  \label{fig:c_par1}
\end{subfigure}
\begin{subfigure}{.24\textwidth}
  \centering
  \includegraphics[width=4.0cm]{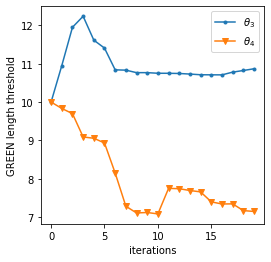}  
  \caption{}
  \label{fig:c_par2}
\end{subfigure}
\begin{subfigure}{.24\textwidth}
  \centering
  \includegraphics[width=4.0cm]{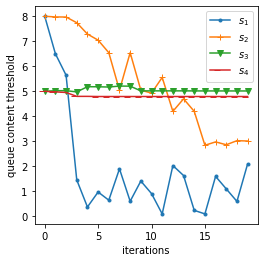}  
  \caption{}
  \label{fig:c_par3}
\end{subfigure}
\caption{Sample cost and parameter trajectories for $1/\bar{\alpha}=[6,6,10,20]$}
\label{fig:c_all}
\end{figure*}

\textbf{Optimizing cost.} 
We use the same initial parameter values $\upsilon_0=[10,20,30,50,10,10,8,8,5,5]$ over different traffic conditions (indicated by different Poisson rates) to test how the controller performs. The optimal average waiting time is recorded after 20 times of parameter ($\upsilon$) updates. 
The direction of each update is based on the average gradient, calculated by IPA using 20 sample paths with a length of 1000s each. 
The results are shown in Table \ref{table1}. For different traffic conditions(denoted by $1/\bar{\alpha}$) in the first column, corresponding weighted average waiting times using the same initial parameters $\upsilon_0$ are recorded in column $J_{init}$. Then, after 20 iterations of parameter adjustment, we record the near-optimal performances and their corresponding parameter values in column $J_{opt}$ and $\upsilon_{opt}$ respectively. 
The reduction of waiting time in the last column shows that our TLC method can improve performance under different traffic intensities, with waiting time reduction varying from 33.8\% to 62.9\%. 
Figure \ref{fig:c_performance} shows the cost trajectory for $1/\bar{\alpha}=[6,6,10,20]$ as the number of parameter iterations increases. Note that since the vehicle flow is higher than the pedestrian flow, the overall weighted mean waiting time curve is dominated by the former. 
Figures \ref{fig:c_par1},\ref{fig:c_par2} and \ref{fig:c_par3} show the convergence of the 10 controllable parameters indicating that the TLC policies designed
in section \ref{section:Controller Specification} are robust to small parameter changes and can produce improved, as well as stable, performance.

\begin{table*}
\caption{Simulation result for different traffic intensity}
\label{table1}
\begin{center}
\resizebox{.7\textwidth}{!}{
\begin{tabular}{|c||c||c||c||c|}
\hline
$1/\bar{\alpha}$                       & $J_{init}$ & $J_{opt}$ & $\upsilon_{opt}$                                                   & Cost Reduction \\ \hline
{[}5,5,20,20{]}                        & 19.08      & 12.63    & {[}11.04, 25.28, 8.52, 49.69, 14.93, 6.06, 4.08, 6.94, 5.0, 4.86{]}              & 33.8\%         \\ \hline
{[}5,6,20,20{]}                        & 15.74     & 7.42     & {[}8.40, 24.97, 0.79, 49.64, 15.50, 0.10, 3.06, 2.43, 4.90, 4.87{]}    & 52.8\%         \\ \hline
 {[}5,7,20,20{]} & 15.32      & 6.51      & {[}6.60, 20.92, 2.72, 49.66, 13.01, 0.10, 1.60, 2.49, 4.97, 4.58{]}   & 57.5\%         \\ \hline
{[}5,8,20,20{]}                        & 12.58      & 4.67      & {[}9.46, 12.10, 3.75, 49.49, 12.55, 2.82, 2.75, 9.63, 5.21, 4.94{]} & 62.9\%         \\ \hline
{[}6,6,20,20{]}                        & 11.11      & 5.68      & {[}9.13, 12.45, 7.39, 48.55, 9.46, 8.01, 0.60, 3.14, 4.96, 4.71{]}  & 48.9\%         \\ \hline
{[}6,7,20,20{]}                        & 10.05      & 4.26      & {[}9.11, 9.11, 0.27, 48.59, 7.67, 9.23, 1.91, 7.75, 4.96, 4.20{]}   & 57.6\%         \\ \hline
{[}6,8,20,20{]}                        & 9.45       & 3.58      & {[}9.70, 10.19, 2.67, 48.91, 7.96, 8.49, 1.60, 7.66, 4.96, 4.83{]}   & 62.1\%         \\ \hline
{[}7,7,20,20{]}                        & 7.75       & 4.00         & {[}9.91, 9.91, 8.18, 48.09, 7.69, 7.29, 0.10, 9.27, 4.99, 4.52{]}   & 48.4\%         \\ \hline
{[}7,8,20,20{]}                        & 7.82       & 3.29      & {[}9.66, 9.66, 2.89, 48.25, 7.04, 7.56, 1.60, 7.73, 5.0, 4.19{]}    & 57.9\%         \\ \hline
{[}8,8,20,20{]}                        & 6.98       & 3.01      & {[}9.91, 9.91, 7.96, 47.63, 7.27, 7.29, 0.10, 7.93, 4.99, 4.27{]}   & 56.9\%         \\ \hline
 {[}6,6,10,20{]} & 11.72      & 6.19      & {[}8.55, 14.96, 6.50, 49.31, 10.71, 7.40, 0.10, 2.84, 5.01, 4.79{]}   & 47.2\%         \\ \hline
{[}6,6,15,20{]}                        & 10.98      & 5.08      & {[}9.75, 10.21, 5.63, 48.73, 8.85, 8.90, 2.10, 7.31, 4.96, 4.67{]}   & 53.7\%         \\ \hline
{[}6,6,25,20{]}                        & 11.50      & 5.00         & {[}6.29, 10.27, 3.66, 48.88, 10.20, 8.49, 1.10, 3.02, 4.96, 4.71{]}  & 56.5\%         \\ \hline
\end{tabular}%
}
\end{center}
\end{table*}





\textbf{TLC of an actual intersection.}
We have cooperated with the town of Veberöd in Sweden to study a major intersection at the town center. This is a typical 4-way intersection with a single lane for each direction. Currently, no traffic light is present and vehicles and pedestrians follow the first-come-first-leave rule for crossing. Since pedestrians always have the right of way, we can regard the operation of this intersection as employing an always-green traffic light except when a pedestrian asks for crossing. When traffic is sparse, such ``random crossing'' is smooth and efficient. However, during busy hours, this rule can cause long queues and congestion. Moreover, is is not safe for crossing pedestrians who need to find a gap through crowded traffic. Furthermore, traffic is expected to increase in the near future as the town expands and develops. 

\begin{figure}
\centering
\includegraphics[width=7.5cm]{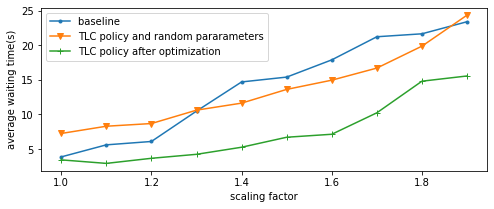}
\caption{\label{Veb_res}Comparison of performance measure}
\end{figure}

With Veberöd traffic data available, we have abstracted the arrival processes to be Poisson with rates $\bar{\alpha}=[0.11,0.125,0.01, 0.01]$. 
We first simulate the intersection operation under current conditions to establish a baseline, and then compare the result to operation using our controller. 
Then, we proportionally expand the traffic applying increasing scaling factors, to imitate anticipated future town development.
The results are shown in Figure \ref{Veb_res}, comparing the baseline (no control) to our adaptive quasi-dynamic controller operating with random unoptimized initial parameters, as well as with with those optimized by the IPA gradient-based method.
We can see the benefits of the TLC which results in cost decreases varying from 11.13\% to 64.22\%. It is also worth noting that when the scaling factor is small (less than 1.3), the current baseline policy works better than the unoptimized traffic light policy. 

\begin{figure*}[ht]
\begin{subfigure}{.24\textwidth}
  \centering
  \includegraphics[width=4cm]{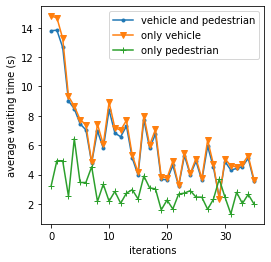} 
  \caption{}
  \label{fig:sequential_performance}
\end{subfigure}
\begin{subfigure}{.24\textwidth}
  \centering
  \includegraphics[width=4.0cm]{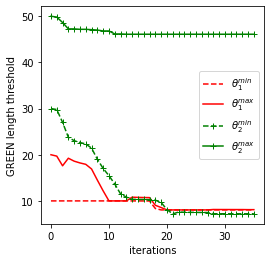}  
  \caption{}
  \label{fig:sequential_par1}
\end{subfigure}
\begin{subfigure}{.24\textwidth}
  \centering
  \includegraphics[width=4.0cm]{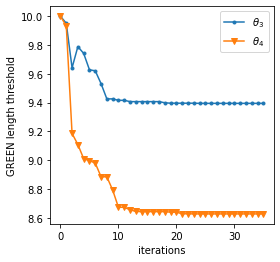}  
  \caption{}
  \label{fig:sequential_par2}
\end{subfigure}
\begin{subfigure}{.24\textwidth}
  \centering
  \includegraphics[width=4.0cm]{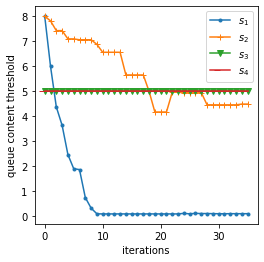}  
  \caption{}
  \label{fig:sequential_par3}
\end{subfigure}
\caption{Sample cost and parameter trajectories for $\bar{\alpha}=[0.154, 0.175, 0.014, 0.014]$}
\label{fig:sequential_all}
\end{figure*}

\begin{figure}
\centering
\includegraphics[width=7.5cm]{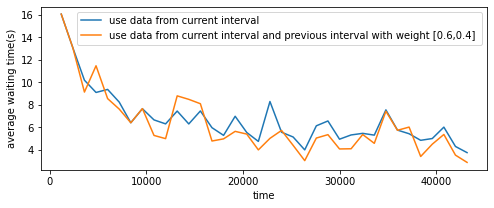}
\caption{\label{fig:sequential_compare}Comparison of using different data}
\end{figure}

\textbf{Online TLC implementation.} 
The ultimate goal of TLC is to operate on line, i.e., observe real-time traffic data and adaptively adjust the controllable parameters. 
We simulate this process by creating a single long sample path. With every observed event occurrence (assuming some sensing capabilities)
IPA derivative updates are triggered accordingly, the results are accumulated, and the parameters are updated periodically. 
We set the Poisson traffic demand at rates $\bar{\alpha}=[0.154, 0.175, 0.014, 0.014]$ ($1.4$ times of Veberöd traffic demand in last case), and the same initial parameter as before. 
The sample path length is $T=43200 s$, and parameters are updated every $1200 s$ using the data collected during the most recent time window. Typical sample trajectories of the cost and parameter changes are shown in Fig.\ref{fig:sequential_all}. 
Observe that cost trajectory fluctuations occur even after parameters have largely converged, indicating, as expected, that the cost is subject to the noise due to the random traffic.
In order to decrease such fluctuations caused by both traffic demand changes and parameter updates, several smoothing techniques can be applied, such as making use of data from prior estimation intervals along with data from the current interval using adjustable weights that emphasize recent data. For example, combining the data from last two time intervals with weight $0.6$ and $0.4$ for most recent data and data from previous interval respectively shows good performance in terms of decreasing the cost variance as shown in Fig.\ref{fig:sequential_compare}, which indicates that when facing the same traffic scenario, cost by using only data from current interval shows higher variance than using combination of data after converging.

\begin{figure}
\centering
\includegraphics[width=7.5cm]{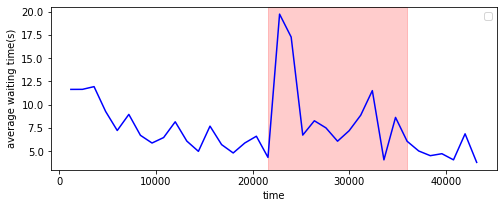}
\caption{\label{fig:adaptivity}Sample cost trajectory with traffic demand perturbation}
\end{figure}

\textbf{TLC adaptivity.}
Our TLC is designed to adapt to changing traffic conditions.
We illustrate this property by observing how the TLC the performance changes when traffic demand is perturbed. 
Using the same simulation setting as before, we add traffic perturbations by increasing the Poisson rate of vehicle flow $1$ to $1.3$ times at $21600 s$, and then return to the original rate at $36000 s$. The cost trajectory is shown in Fig.\ref{fig:adaptivity} where the shaded area corresponds to the time interval over which traffic demand was increased. 
We can see that the waiting time initially decreases due to our parameters being adjusted as before. However, when traffic demand abruptly increases, the waiting time increases since the previously optimized parameters no longer apply to the new traffic demand. Nonetheless they immediately adjust and converge to new optimal values after several iterations. 

\section{Conclusion and Future Work}
We have studied a TLC problem for a single intersection with both vehicle and pedestrian flows. We used a SFM to model this as a stochastic dynamic hybrid system.
We then designed an adaptive quasi-dynamic controller aiming at optimizing the performance (weighted mean waiting time in our case) of this intersection. Our TLC is parameterized and IPA is used to estimate the cost gradient with respect to these parameters applied to both vehicles and pedestrians. We then adjust the parameters iteratively through an online gradient-based algorithm in order to improve overall performance and enable it to automatically adapt to changing traffic conditions. Our next steps are to add more flows, e.g., left-turn and right-turn traffic and bicycle flows. With improving and cost-effective data detection and transmission techniques, we can also expand our TLC to be fully dynamic, thus making the best use of all the information collected. Moreover, we plan to extend the use of such TLCs research to multiple intersections which we expect to cooperate and exhibit ``green wave'' behaviors which are highly desirable.

\bibliographystyle{IEEEtran}
\bibliography{ref}

\appendix
\section{Derivation of event time derivatives} \label{event time derivative analysis}
In this section, we derive the event time derivatives with respect to each of the controllable parameters $\upsilon$ formulated in (\ref{parametervector}). We use $\upsilon_i$ as $i$th parameter(e.g., $\upsilon_1=\theta_1^{min}$), and $i$ indicates the index of parameter through whole section. So that we define $\tau_{k,i}'$ as the derivative of $k$th event time $\tau_k$ with respect to $i$th parameter, which is the target of derivation, for $i=1,...10$.
We take $n=1$ and $u_1(t)=1$ in the following analysis, and since the SHA is symmetric, it's easy to see that situation $u_1(t)=0$ ($u_2(t)=1$) shares the same logic. Note that we are only interested in events which would cause light switching($G2R_1$) in scenarios 3.1, 5.1, and 5.2, and event [$x_1 \downarrow 0$] in scenario 2.
\begin{itemize}
    \item[1)] Event [$\alpha_1\downarrow 0$] occurs at time $\tau_k$ and results in $G2R_1$. Even though it induces light switch, it's an exogenous event so that
      \begin{equation}
        \tau_{k,i}'= 0
    \end{equation}
    
    \item[2)] Event [$z_1\uparrow \theta_1^{max}$] occurs at time $\tau_k$ and result in $G2R_1$: The guard condition is $g_k=z_1-\theta_1^{max}$, where the index of $\theta_1^{max}$ is $2$ in parameter vector (\ref{parametervector}), so that $\theta_1^{max}=\upsilon_2$ . We have $\frac{\partial g_k}{\partial z_1}=1$ and $\frac{\partial g_k}{\partial \upsilon_2}=-1$, and all other partial derivatives equals to zero. Since it's GREEN phase before $\tau_k$, we have $f_{1,k-1}^z(\tau_k^-)=1$ and $f_{1,k}^z(\tau_k^+)=0$. Then based on (\ref{eqn:event_time_derivative}), it remains to find $z'_{1,i}(\tau_k^-)$. We define $\tau_{k-1}$ as the event time of last light switch ($R2G_1$) and $\tau_{k-2}$ be the light switch event time before that($G2R_1$) as shown in Fig.\ref{fig:z1_tauk}. With (\ref{eqn:derivative within mode}), we can know that $z'_{1,i}(\tau_k^-)=z'_{1,i}(\tau_{k-1}^+)$, and $z'_{1,i}(\tau_{k-1}^-)=z'_{1,i}(\tau_{k-2}^+)$. Also, since $z_1$ would be reset to zero when event $G2R_1$ happens, so that $z'_{1,i}(\tau_{k-2}^+) =0= z'_{1,i}(\tau_{k-1}^-)$. And based on (\ref{eqn:state_derivative}), $z'_{1,i}(\tau_{k-1}^+)=-\tau_{k-1,i}' = z'_{1,i}(\tau_{k}^-) $. Substitute it in (\ref{eqn:event_time_derivative}), we get:
    \begin{equation}\label{eqn:event_der_z1_theta1max}
        \tau_{k,i}'= \tau_{k-1,i}'+\mathds{1}_{i=2}
    \end{equation}
     where $\mathds{1}_{i=2}$ is an indicator function which equals to $1$ when condition $i=2$ satisfies. Similar notation applies in the following content. Equation (\ref{eqn:event_der_z1_theta1max}) means the event time derivative would inherit the old value only except $i=2$ when another $1$ would be added.

\begin{figure}
\centering
\includegraphics[width=7.5cm]{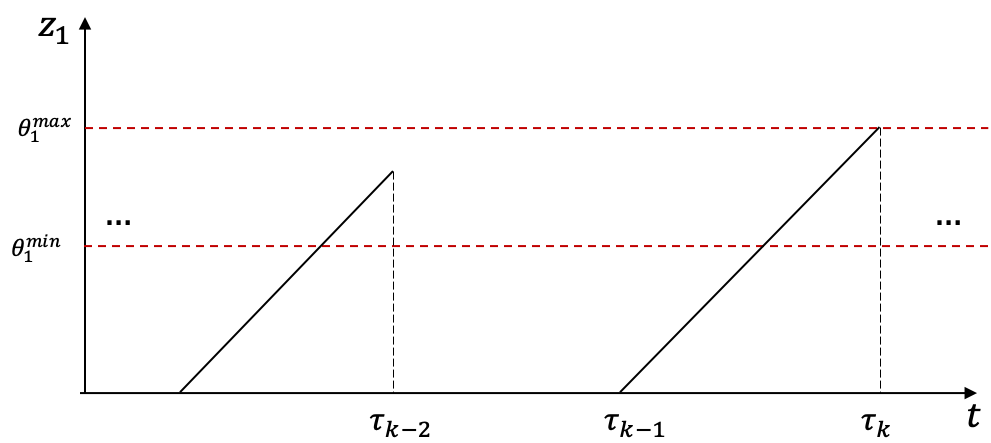}
\caption{\label{fig:z1_tauk}A sample path of $z_1$ }
\end{figure}

   \item[3)] Event [$z_1\uparrow \theta_1^{min}$] occurs at time $\tau_k$ and result in $G2R_1$.
   The guard condition is $g_k=z_1-\theta_1^{min}=0$. Similar as last condition, we have $\frac{\partial g_k}{\partial z_1}=1$ and $\frac{\partial g_k}{\partial \upsilon_1}=-1$, and
    \begin{equation}
        \tau_{k,i}'= \tau_{k-1,i}'+\mathds{1}_{i=1}
    \end{equation}

    \item[4)] Event [$x_1\downarrow s_1$] occurs at time $\tau_k$ and result in $G2R_1$. This is an endogenous event with the guard condition $g_k=x_1-s_1$. And we then have $\frac{\partial g_k}{\partial x_1}=1$, $\frac{\partial g_k}{\partial \upsilon_7}=-1$, and all other partial derivatives equals to zero. Due to the light switching, the dynamic changes from $f_{1,k-1}^x(\tau_k^-)=\alpha_1(\tau_k)-h_1(\tau_k)$ to $f_{1,k}^x(\tau_k^+)=\alpha_1(\tau_k)$. So that based on (\ref{eqn:event_time_derivative}), the event time derivative is as:
    \begin{equation}\label{eqn:event_der_x1_s1}
        \tau'_{k,i}=\frac{\mathds{1}_{i=7}-x'_{1,i}(\tau_k^-)}{\alpha_1(\tau_k)-h_1(\tau_k)}
    \end{equation}

    \item[5)] Event [$x_2\uparrow s_2$] occurs at time $\tau_k$ and result in $G2R_1$. This is an endogenous event with similar guard condition $g_k=x_2-s_2$. So we have $\frac{\partial g_k}{\partial x_2}=1$, $\frac{\partial g_k}{\partial \upsilon_8}=-1$, and all other partial derivatives equals to zero. And since $f_{2,k-1}^x(\tau_k^-)=\alpha_2(\tau_k)$ and $f_{2,k}^x(\tau_k^+)=\alpha_2(\tau_k)-h_2(\tau_k)$ as in (\ref{eqn:xdynamic}), using (\ref{eqn:event_time_derivative}), we have
    \begin{equation}\label{eqn:event_der_x2_s2}
        \tau'_{k,i}=\frac{\mathds{1}_{i=8}-x'_{2,i}(\tau_k^-)}{\alpha_2(\tau_k)}
    \end{equation}

    \item[6)] Event [$x_3\uparrow s_3$] occurs at time $\tau_k$ and result in $p_1=1$ and $G2R_1$. This is an endogenous event with guard condition $g_k=x_3-s_3$. With a similar analysis as (\ref{eqn:event_der_x2_s2}), we have:
    \begin{equation}\label{eqn:event_der_x3_s3}
         \tau'_{k,i}=\frac{\mathds{1}_{i=9}-x'_{3,i}(\tau_k^-)}{\alpha_3(\tau_k)}
    \end{equation}

    \item[7)] Event [$w_3\uparrow \theta_3$] occurs at time $\tau_k$ and result in $p_1=1$ and $G2R_1$. Also, it's an endogenous event with guard condition $g_k=w_3-\theta_3$. So that we have $\frac{\partial g_k}{\partial w_3}=1$, $\frac{\partial g_k}{\partial \upsilon_5}=-1$ and all other partial derivatives equal to $0$. The dynamic of state $w_3$ changes from $f_{3,k-1}^w(\tau_k^-)=1$ to $f_{3,k}^w(\tau_k^+)=0$, which lead to $\tau'_{k,i}=\mathds{1}_{i=5}-w_{3,i}'(\tau_k^-)$ according to (\ref{eqn:event_time_derivative}). In order to find $w_{3,i}'(\tau_k^-)$, we define $\tau_{k-1}$ as the event time of last light switch ($R2G_1$) and $\tau_{k-2}$ be the light switch event time before that($G2R_1$). Also, we define $\sigma_k<\tau_k$ as the event time of [$\alpha_3\uparrow 0$] in current phase, when $w_3$ changes from $0$ to positive. Similar to previous analysis for (\ref{eqn:event_der_z1_theta1max}), it's obvious that $w'_{3,i}(\sigma_k^-)=w'_{3,i}(\tau_{k-2}^+)=0$. So that $w'_{3,i}(\tau_k^-)=w'_{3,i}(\sigma_k^+)=w'_{3,i}(\sigma_k^-) + [f_{3,k-1}^w(\sigma_k^-)-f_{3,k}^z(\sigma_k^+)]\sigma_k' =-\sigma'_k$, according to (\ref{eqn:state_derivative}). Also, since $\sigma_k$ is the time of event [$\alpha_3\uparrow 0$], which is an exogenous event, $\sigma'_k=0$. Combine all of these, we have
    \begin{equation}
         \tau'_{k,i}=\mathds{1}_{i=5}
    \end{equation}

    \item[8)] Event [$x_4\downarrow s_4$] occurs at time $\tau_k$ and result in $p_2=0$ and $G2R_1$. This is an endogenous event with guard condition $g_j=x_4-s_4$. The state dynamic changes from $f_{4,j}^x(\tau_j^-)=\alpha_4(\tau_j)-h_4(\tau_k)$ to $f_{4,j}^x(\tau_j^-)=\alpha_4(\tau_j)$. Similar to analysis for (\ref{eqn:event_der_x1_s1}), we have:
    \begin{equation}
         \tau'_{j,i}=\frac{\mathds{1}_{i=10}-x'_{4,i}(\tau_j^-)}{\alpha_4(\tau_j)-h_4(\tau_k)}
    \end{equation}

    \item[9)] Event [$x_1 \downarrow 0$] occurs at time $\tau_k$. The guard condition is $g_k=x_1-0=0$, which gives $\frac{\partial g_k}{\partial x_1}=1$, $\frac{\partial g_k}{\partial \upsilon_i}=0$. And the dynamics are $f_{1,k-1}^x(\tau_k^-)=\alpha_1(\tau_k)-h_1(\tau_k)$, $f_{1,k}^x(\tau_k^+)=\alpha_1(\tau_k)-\alpha_1(\tau_k)=0$. Then, we can derive from (\ref{eqn:event_time_derivative}) that:
    \begin{equation}
    \tau_{k,i}'= \frac{-x_{1,i}'(\tau_k^-)}{\alpha_1(\tau_k)-h_1(\tau_k)} 
    \end{equation}

\end{itemize}

\end{document}